\documentclass[useAMS,usenatbib]{mn2e}

\usepackage{natbib}
\usepackage{graphicx}
\usepackage{amssymb}
\usepackage{color}
\usepackage{caption}

\title[AGN-SB connection]{AGN-starburst evolutionary connection : a physical interpretation based on radiative feedback}

\author[ ]
{W. Ishibashi$^{1}$\thanks{E-mail:
wako.ishibashi@phys.ethz.ch} and A. C. Fabian$^{2}$
\footnotemark[0]\\
$^{1}$Institute for Astronomy, Department of Physics, ETH Zurich, Wolfgang-Pauli-Strasse 27, CH-8093 Zurich, Switzerland 
\footnotemark[0]\\
$^{2}$Institute of Astronomy, Madingley Road, Cambridge CB3 0HA 
}

\begin{document}

\pdfminorversion=4

\date{Accepted ? Received ?; in original form ? }


\pagerange{\pageref{firstpage}--\pageref{lastpage}} \pubyear{2012}

\maketitle

\label{firstpage}

\begin{abstract} 
Observations point towards a close connection between nuclear starbursts, active galactic nuclei (AGN), and outflow phenomena. An evolutionary sequence, starting from a dust-obscured ultra-luminous infrared galaxy and eventually leading to an unobscured optical quasar, has been proposed and discussed in the literature. AGN feedback is usually invoked to expel the obscuring gas and dust in a blow-out event, but the underlying physical mechanism remains unclear. We consider AGN feedback driven by radiation pressure on dust, which directly acts on the obscuring dusty gas. We obtain that radiative feedback can potentially disrupt dense gas in the infrared-optically thick regime, and that an increase in the dust-to-gas fraction leads to an increase in the effective Eddington ratio. Thus the more dusty gas is preferentially expelled by radiative feedback, and the central AGN is prone to efficiently remove its own obscuring dust cocoon. Large amounts of dust imply heavy obscuration but also powerful feedback, suggesting a causal link between dust obscuration and blow-out. In this picture, AGN feedback and starburst phenomena are intrinsically coupled through the production of dust in supernova explosions, leading to a natural interpretation of the observed evolutionary path. 
\end{abstract}



\section{Introduction}

Observations indicate close couplings between star formation in nuclear starbursts and the growth of super-massive black holes in active galactic nuclei (AGN) \citep[e.g.][and references therein]{Alexander_Hickox_2012}. 
In co-evolution scenarios, galaxy mergers or other secular processes are believed to funnel large quantities of gas into the central regions. Part of the gas fuels the nuclear starburst while some gas may leak in to feed the central black hole, the whole process being governed by angular momentum transfer. Large amounts of matter are potentially available for accretion, but that same accreting material also provides significant obscuration. 

An evolutionary sequence linking ultra-luminous infrared galaxies (ULIRG) to luminous optical quasars (QSO), with a transition through a heavily obscured and dust-enshrouded phase, has been widely discussed in the literature \citep[since][]{Sanders_et_1988}. 
In these scenarios, the dust initially absorbs most of the quasar ultraviolet (UV) radiation, which is then reprocessed and re-emitted as infrared (IR) radiation. At this stage, the central source can be observationally detected as an ULIRG. 
Following the removal of the dust shroud in a short-lived blow-out event, the central source may eventually be observed as an unobscured UV-luminous quasar. 

Many observational works have searched for `transition' objects, i.e. sources evolving from the dust-obscured starburst phase towards the unobscured QSO stage. The search has been extended to high redshifts, close to the peak epoch of both AGN and star formation activities ($z \gtrsim 2$), where submillimetre galaxies (SMG) represent high-redshift analogs of local ULIRG \citep{Coppin_et_2008, Simpson_et_2012}. New populations of dust-reddened quasars, likely observed in the short-lived blow-out phase, have been recently uncovered \citep{Banerji_et_2012, Glikman_et_2012}. The luminous red quasars are characterised by high Eddington ratios, with a significant fraction of the population showing direct signatures of outflowing gas in the form of line broadening and BAL features, supporting the evolutionary picture (cf Sect. \ref{Obs_samples}). 

On the other hand, numerical simulations have tried to reproduce the temporal evolution of the proposed sequence, starting from major mergers of gas-rich galaxies and including AGN feedback \citep[e.g.][]{diMatteo_et_2005, Hopkins_et_2005}. The feedback process is usually implemented by coupling a fixed fraction of the accretion luminosity to the surrounding medium. The feedback energy is assumed to heat the ambient gas, driving a powerful wind that sweeps the obscuring gas and dust in a blow-out event. 
However, the AGN feedback mechanism is not specified, and in particular there is no explicit connection with the dust component.

Here we wish to consider a physical mechanism that directly makes use of the dust in order to drive AGN feedback.  
The effects of radiation pressure on dusty gas and the potential role in driving AGN feedback have been previously considered \citep{Fabian_1999, Murray_et_2005}. 
Possible indications of the existence of an effective Eddington limit determined by radiation pressure on dust are found in X-ray surveys, as a `forbidden zone' observed in the $N_H - \lambda$ plane \citep[e.g.][]{Fabian_et_2008}. 
Numerical simulations of AGN feedback including radiation pressure on dust have also been performed \citep{Debuhr_et_2011, Roth_et_2012}. 
\citet{Thompson_et_2015} have recently analysed the dynamics of dusty shells driven by radiation pressure, showing that the shell asymptotic velocity can be much higher than the local escape velocity in different astrophysical sources. 
Focusing on AGN feedback, we have further studied the role of radiation pressure on dust in driving powerful outflows on galactic scales \citep{Ishibashi_Fabian_2015}.  
Below we briefly discuss how AGN radiative feedback may provide a physical explanation for the removal of obscuring dusty gas, leading to a natural interpretation of the observed evolutionary path.


\section{Effective Eddington ratios}

We consider radiation pressure on dusty gas sweeping up ambient material into an outflowing shell. 
The general form of the equation of motion is given by 
\begin{equation}
\frac{d}{dt} [M_{sh}(r) v] = \frac{L}{c} (1 + \tau_{IR} - e^{-\tau_{UV}}) - \frac{G M(r) M_{sh}(r)}{r^2}
\end{equation} 
where $L$ is the central luminosity, $M(r)$ the total mass distribution, and $M_{sh}(r)$ the shell mass \citep[cf][]{Thompson_et_2015, Ishibashi_Fabian_2015}. 
The associated shell column density is defined as:
\begin{equation}
N_{sh}(r) = \frac{M_{sh}(r)}{4 \pi r^2 m_p} 
\end{equation}
The IR and UV optical depths are respectively given by
\begin{equation}
\tau_{IR}(r) = \frac{\kappa_{IR} M_{sh}(r)}{4 \pi r^2} = \kappa_{IR} m_p N_{sh}(r)
\end{equation}
\begin{equation}
\tau_{UV}(r) = \frac{\kappa_{UV} M_{sh}(r)}{4 \pi r^2} = \kappa_{UV} m_p N_{sh}(r)
\end{equation}
where $\kappa_{IR}$=$5 cm^2 g^{-1} f_{dg, MW}$ and $\kappa_{UV}$=$10^3 cm^2 g^{-1} f_{dg, MW}$ are the IR and UV opacities, with the dust-to-gas ratio normalised to the Milky Way value. 

A critical luminosity is obtained by equating the outward force due to radiation pressure to the inward force due to gravity, which can be considered as a generalised form of the Eddington luminosity ($L_E'$). 
Dusty gas is ejected when the central luminosity exceeds the effective Eddington limit \citep[cf][]{Ishibashi_Fabian_2016}.  
The corresponding Eddington ratio can be defined as:
\begin{equation}
\Gamma = \frac{L}{L_E'} = \frac{L r^2}{c G M(r) M_{sh}(r)} (1 + \tau_{IR} - e^{-\tau_{UV}}) 
\end{equation}

We note that the Eddington ratio in its generalised form basically corresponds to the ratio of the radiative force to the gravitational force ($\Gamma = \frac{F_{rad}}{F_{grav}}$). 
In the single scattering regime, the Eddington ratio is given by 
\begin{equation}
\Gamma_{SS} = \frac{L}{4 \pi G c m_p M(r) N_{sh}(r)}  
\label{Gamma_SS}
\end{equation}
while the Eddington ratios in the IR-optically thick and UV-optically thin regimes are given by
\begin{equation}
\Gamma_{IR} =  \frac{\kappa_{IR} L}{4 \pi G c M(r)} 
\label{Gamma_IR}
\end{equation}
\begin{equation}
\Gamma_{UV} = \frac{\kappa_{UV} L}{4 \pi G c M(r)} 
\label{Gamma_UV}
\end{equation}

We see that in the single scattering regime, the Eddington ratio scales inversely with the column density and is independent of the opacities of the ambient medium. In contrast, in the IR-optically thick and UV-optically thin regimes, the Eddington ratio is independent of the column density, while it directly scales with the IR and UV opacity, respectively. In the single scattering regime, the Eddington ratio decreases for increasing columns, and gravity tends to dominate over the radiative force. However, as the IR optical depth becomes large enough, the system enters in the IR-optically thick regime where the Eddington ratio is independent of $N_{sh}$. In this regime, the Eddington ratio is not further reduced for increasing columns. 

\begin{figure}
\begin{center}
\includegraphics[angle=0,width=0.4\textwidth]{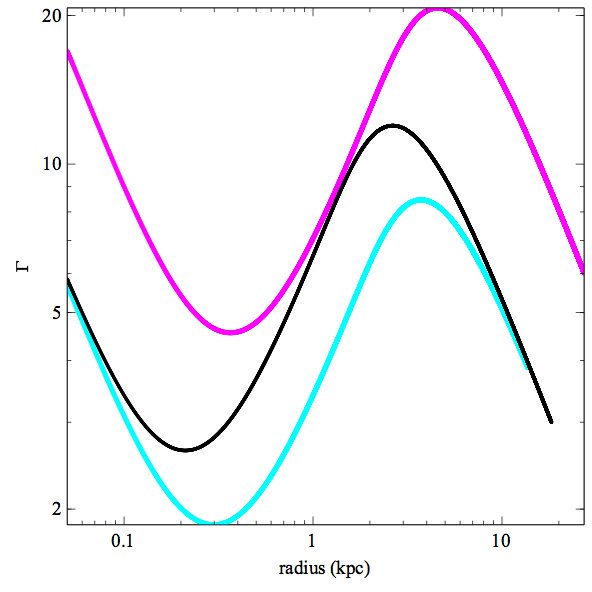} 
\caption{\small
Effective Eddington ratio as a function of radius:
$L = 5 \times 10^{46}$erg/s, $M_{sh} = 5 \times 10^8 M_{\odot}$, $f_{dg} = 1/150$ (black); $L = 5 \times 10^{46}$erg/s, $M_{sh} = 1 \times 10^9 M_{\odot}$, $f_{dg} = 1/150$ (cyan); $L = 5 \times 10^{46}$erg/s, $M_{sh} = 5 \times 10^8 M_{\odot}$, $f_{dg} = 1/50$ (magenta).
}
\label{Fig_Eddington_ratios}
\end{center}
\end{figure} 

We illustrate these points in Figure \ref{Fig_Eddington_ratios}, where we plot the effective Eddington ratio as a function of radius in the simple case of an isothermal potential ($M(r) = \frac{2 \sigma^2}{G} r$, where $\sigma$ is the velocity dispersion) and fixed-mass shell ($M_{sh}(r) = M_{sh}$). 
We have previously considered the dynamics of expanding shells sweeping up matter from the surrounding medium, showing that such shells likely remain bound on large scales and possibly fall back \citep{Ishibashi_Fabian_2015, Ishibashi_Fabian_2016}. Here we are interested in the blow-out phase and focus on fixed-mass shells, also noting that the shell mass configuration is irrelevant in the IR-optically thick regime. 
The Eddington ratios in the three regimes are respectively given by:
\begin{equation}
\Gamma_{SS} = \frac{L}{2 c \sigma^2 M_{sh}} r  
\end{equation}
\begin{equation}
\Gamma_{IR} = \frac{\kappa_{IR} L}{8 \pi c \sigma^2 r} 
\end{equation}
\begin{equation}
\Gamma_{UV} = \frac{\kappa_{UV} L}{8 \pi c \sigma^2 r} 
\end{equation} 

As expected, we see that an increase in the shell mass leads to a lower Eddington ratio in the single scattering regime (cyan curve in Figure \ref{Fig_Eddington_ratios}). In contrast, the Eddington ratio in the IR-optically thick and UV-optically thin regimes are independent of the shell mass, and thus the cyan and black curves overlap at small and large radii. 
On the other hand, enhanced opacities (e.g. due to higher dust-to-gas ratios) lead to higher Eddington ratios in the IR-optically thick and UV-optically thin regimes (magenta curve in Figure \ref{Fig_Eddington_ratios}). But in the single scattering regime, $\Gamma_{SS}$ is independent of the medium opacities, and thus the magenta and black curves overlap at intermediate radii. 

Dusty gas surrounding the central source absorbs UV radiation and re-emits in the IR band. 
If the reprocessed IR photons remain trapped, the system is effectively in the IR-optically thick domain. 
In this regime, the effective Eddington ratio is independent of the column density, implying that even dense material can potentially be disrupted. 
Such conditions of high IR optical depth may be commonly reached in the nuclear regions of ULIRG-like systems, characterised by high densities and high dust-to-gas fractions. 
Indeed, starburst galaxies are characterised by high star formation rates, and presumably high supernova rates, which effectively contribute to dust production (cf Section \ref{Sect_Discussion}). 
Large values of the dust-to-gas ratio ($f_{dg} \sim 1/50$) are observed in dense starbursts, such as SMG at high redshifts \citep{Kovacs_et_2006, Michalowski_et_2010}. The dust-to-gas ratio is also known to correlate with metallicity and radius, with $f_{dg}$ reaching higher values in the inner regions \citep[][and references therein]{Andrews_Thompson_2011}. 
As the IR opacity directly scales with the dust-to-gas ratio, enhanced $f_{dg}$ fractions lead to both higher IR optical depth and higher IR-Eddington ratio, which combine to facilitate the blow-out of dusty gas.


\section{Obscuration and momentum ratio}

\begin{figure}
\begin{center}
\includegraphics[angle=0,width=0.4\textwidth]{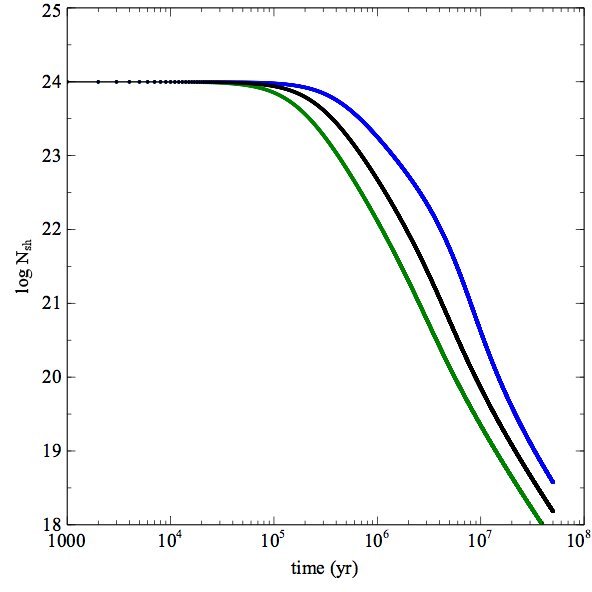} 
\caption{\small Column density as a function of time for variations in luminosity ($N_{sh,0} = 10^{24} cm^{-2}$, $f_{dg} = 1/150$): $L = 3 \times 10^{46}$erg/s (blue), $L = 5 \times 10^{46}$erg/s (black), $L = 1 \times 10^{47}$erg/s (green). 
}
\label{Fig_Nsh_t}
\end{center}
\end{figure}

\begin{figure}
\begin{center}
\includegraphics[angle=0,width=0.4\textwidth]{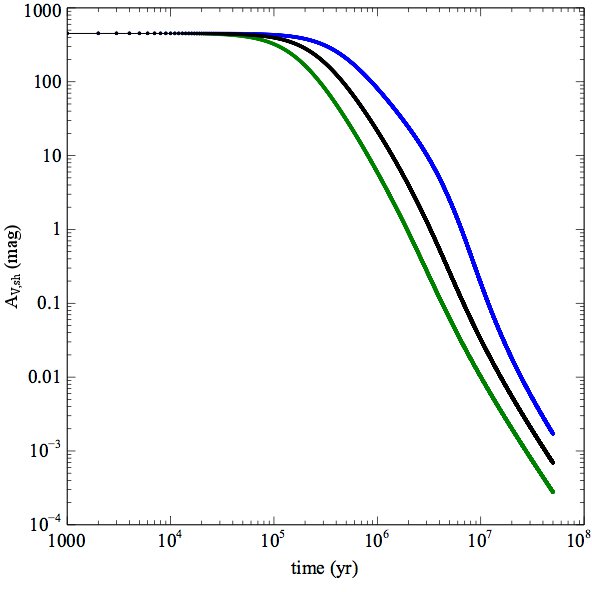} 
\caption{\small Extinction as a function of time for variations in luminosity ($N_{sh,0} = 10^{24} cm^{-2}$, $f_{dg} = 1/150$): $L = 3 \times 10^{46}$erg/s (blue), $L = 5 \times 10^{46}$erg/s (black), $L = 1 \times 10^{47}$erg/s (green). 
}
\label{Fig_AVsh_t}
\end{center}
\end{figure}

The large amount of gas and dust surrounding the central source and forming the absorbing medium is also responsible for significant obscuration. 
Figure \ref{Fig_Nsh_t} shows the temporal evolution of the shell column density for different values of the central luminosity. 
We see that the column density decreases with time, and that the decline is more rapid for brighter sources: at a given time, the obscuring column is lower for higher luminosity. In fact, from Eqs. \ref{Gamma_SS}-\ref{Gamma_UV}, we see that an increase in luminosity leads to an increase in the effective Eddington ratio in all three regimes. The resulting acceleration is most efficient, the outflowing shell reaches a high velocity, and the obscuration rapidly falls off. 
We can also express the evolution of the shell column density in terms of the extinction (Fig. \ref{Fig_AVsh_t}), by assuming the linear relation between optical extinction and hydrogen column density observed in the Galaxy, $N_H (\mathrm{cm^{-2}}) = (2.21 \pm 0.09) \times 10^{21} A_V (\mathrm{mag})$ \citep{Guver_Ozel_2009}. 

For a given luminosity, the Eddington ratio is higher for lower mass shells in the single scattering regime. 
Physically, lower column density material may be more easily expelled, while higher columns are less easily disrupted and tend to survive against radiation pressure. 
As the column density decreases with time, the effective Eddington ratio increases in the single scattering regime. 
The outflowing shell is thus unstable to radial perturbations, and tends to become increasingly super-Eddington with increasing radius \citep[as discussed in][]{Thompson_et_2015}. 

\begin{figure}
\begin{center}
\includegraphics[angle=0,width=0.4\textwidth]{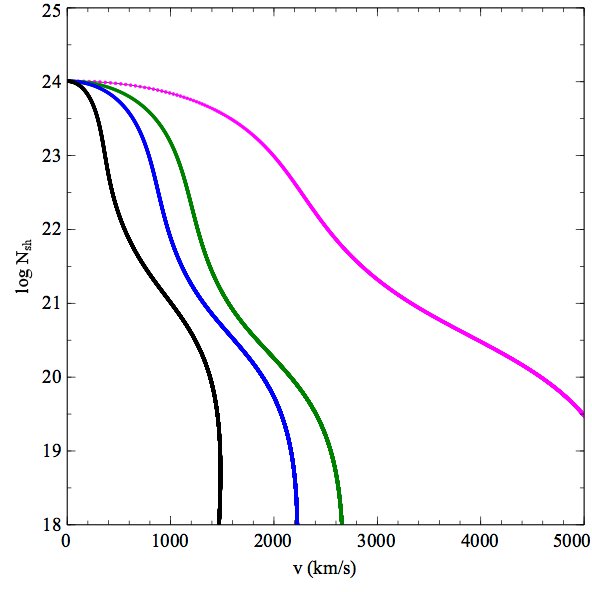} 
\caption{\small Column density as a function of velocity for variations in the dust-to-gas ratio ($L = 2 \times 10^{46}$erg/s, $N_{sh,0} = 10^{24} cm^{-2}$):  $f_{dg} = 1/150$ (black), $f_{dg} = 1/50$ (blue), $f_{dg} = 1/30$ (green). High-luminosity case: $L = 10^{47}$erg/s, $f_{dg} = 1/50$ (magenta). 
}
\label{Fig_Nsh_v_varFdg}
\end{center}
\end{figure}

In Figure \ref{Fig_Nsh_v_varFdg} we plot the shell column density as a function of velocity for different values of the dust-to-gas ratio. 
We observe that a given column is accelerated to higher velocities for larger dust-to-gas fractions: at a given time, the column density is lower for higher dust-to-gas ratios. 
 Indeed, an increase in the dust-to-gas fraction leads to enhanced IR opacity, which in turn leads to a higher Eddington ratio in the IR-optically thick regime. 
Thus the more dusty gas, which also provides much of the obscuration, is more easily ejected by radiative feedback, and the central AGN is prone to efficiently remove its own obscuring dust shroud.

\begin{figure}
\begin{center}
\includegraphics[angle=0,width=0.4\textwidth]{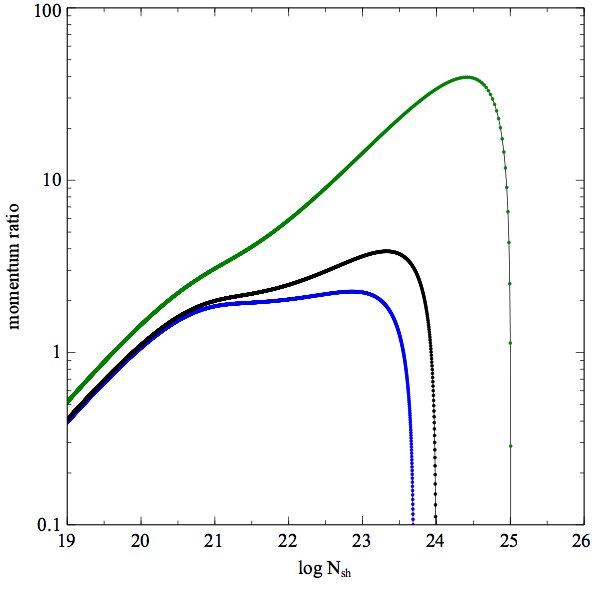} 
\caption{\small Momentum ratio as a function of column density for different values of the initial column ($L = 5 \times 10^{46}$erg/s,  $f_{dg} = 1/150$): $N_{sh,0} = 10^{24} cm^{-2}$ (black), $N_{sh,0} =  5 \times 10^{23} cm^{-2}$ (blue), $N_{sh,0} = 10^{25} cm^{-2}$ (green). 
}
\label{Fig_xi_Nsh_varR0}
\end{center}
\end{figure}

Another important parameter characterising the outflow dynamics is the momentum ratio of the shell, defined as $\zeta = \frac{\dot{M}v}{L/c} = \frac{M_{sh}v^2}{r \frac{L}{c}}$. 
In Figure \ref{Fig_xi_Nsh_varR0}, we plot the momentum ratio as a function of the shell column density. 
We observe that high values of the momentum ratios are associated with high column densities, with the peak $\zeta$ values reached in the highly obscured innermost region. 
We note that large gas masses, of the order of $\sim 10^8 M_{\odot}$, may be involved in the nuclear regions. The presence of such amount of gas may not be implausible in massive quasars, which can double their mass on timescales comparable to the Salpeter time.

Observations of molecular outflows indicate high velocities ($v \gtrsim 1000$km/s), and large momentum flux, with typical values of $(\sim 10)$ $L/c$ \citep{Sturm_et_2011, Veilleux_et_2013, Cicone_et_2014}. 
We have previously shown that large momentum ratios, comparable to the values reported in observations of galactic outflows, can be obtained by taking into account the effects of radiation trapping \citep{Ishibashi_Fabian_2015}. 
Alternatively, we have also suggested that the reported momentum ratios may be over-estimated if the central luminosity has considerably decreased over time due to AGN variability. In such cases, large values of the momentum ratio may be observed at modest columns. 

\section{Observational samples}
\label{Obs_samples}

As discussed in the previous sections, we expect high luminosity sources with significant amount of dust to drive strong outflows via AGN radiative feedback. 
Direct observational evidence of powerful outflows has been recently reported in X-ray obscured QSO at $z \sim 1.5$ \citep{Brusa_et_2015, Perna_et_2015}. These luminous ($L \gtrsim 10^{46}$erg/s), X-ray absorbed ($N_H \sim 10^{21}-10^{22} cm^{-2}$) red QSO show fast outflows with velocities in the range $v \sim 1200-1800$ km/s. From Figure \ref{Fig_Nsh_v_varFdg}, we see that shells with $log N_{sh} \sim 21-22$ can reach velocities of the order of $v \gtrsim 1000$ km/s, depending on the dust-to-gas ratio. For a higher central luminosity (magenta curve), the outflowing shell can reach much higher velocities of several $1000$ km/s, as the effective Eddington ratio increases with luminosity in all three regimes. Such high-luminosity models, coupled with significant amount of dust, may account for the extreme outflows with $v \sim 3000$ km/s observed in red quasars at $z \sim 2.5$ \citep{Zakamska_et_2016}.

Dust-reddened, luminous quasars are likely good candidates for sources in `transition'. 
An unusually high fraction of the red quasar population is found to show BAL characteristics, which is indicative of a blow-out phase when the obscured quasar is expelling its dust cocoon \citep{Urrutia_et_2009, Glikman_et_2012}.
Near-IR selection has unveiled a population of high-redshift ($z \gtrsim 2$), red luminous quasars with significant dust extinction ($A_V \sim 2-6$ mag), which has been interpreted as sources in a short-lived transition phase \citep{Banerji_et_2012, Banerji_et_2015}. 
From Figure \ref{Fig_AVsh_t}, we see that $A_{V,sh}$ of a few would correspond to timescales of $\sim 10^6$yr, consistent with the interpretation that the dust-reddened quasars are observed in a short-lived blow-out stage. 
 Indeed, some of the red quasars show direct evidence for outflows in the form of broadened $H\alpha$ line profiles \citep{Banerji_et_2012}. 
 The next step will be to systematically search for outflow signatures in the dust-reddened quasar population and better characterise their kinematics, e.g. from a detailed analysis of the observed line profiles. 
 
Another class of heavily dust-obscured galaxies, known as Hot DOGs, has been recently identified by WISE \citep[][and references therein]{Assef_et_2015}. Hot DOGs are characterised by high luminosity ($L_{bol} \gtrsim 10^{47}$erg/s) and large obscuration (up to $A_V \sim 50$ mag). Indeed, X-ray observations indicate that Hot DOGs are powered by heavily obscured, possibly Compton-thick AGN \citep{Stern_et_2014, Piconcelli_et_2015}. 
In our picture, high column densities and heavy extinction are expected at early times, with $A_{V,sh} \sim 10-50$ mag corresponding to timescales of a few $\sim 10^5$yr (Figure \ref{Fig_AVsh_t}). 
Such conditions may coincide with the initial stages of an outflow episode. An example is given by the most luminous Hot DOG (W2246-0526), which is observed close to the point of blowing out its dusty cocoon in a large-scale, isotropic outflow \citep{Diaz-Santos_et_2016}. 

The space density of Hot DOGs is found to be comparable to that of equally luminous Type I QSO \citep{Assef_et_2015}; while the 
number density of red quasars with more moderate extinction seems to exceed that of unobscured quasars at the highest luminosities \citep{Banerji_et_2015}. If interpreted as outflowing shells within a temporal sequence, the heavily obscured systems would correspond to an earlier phase preceding the lower extinction stage. In principle, one can try to put constraints on the lifetimes of the obscured and unobscured phases by comparing the number densities of the two populations. However, such comparisons are actually difficult, since Hot DOGs can either have very large black hole masses or accrete at super-Eddington rates \citep{Assef_et_2015}. A more detailed analysis of the relative lifetimes will be required in order to probe any potential evolutionary link between the two populations. 
In any case, a significant population of luminous quasars seems to be dust-obscured, with higher reddenings associated with higher luminosities \citep{Banerji_et_2015}. This suggests a close coupling between extinction and luminosity, as might be expected if the accreting material is also responsible for significant obscuration.


\section{Discussion}
\label{Sect_Discussion}

Although AGN feedback is usually invoked to clear obscuring gas and dust in the blow-out event, the actual physical mechanism remains unclear. In particular, no explicit connection is made with the dust, which plays a crucial role in all the evolutionary scenarios, as it is responsible for limiting the quasar visibility and determining the overall spectral energy distribution. 
In our picture, the driving mechanism for AGN feedback itself is based on radiation pressure on dust.  
Since the same dusty gas, which provides obscuration, forms the medium on which radiative feedback acts, there is a natural causal link between dust obscuration and blow-out. 
Large amounts of dust imply heavy obscuration, but also powerful feedback, and a sort of equilibrium may be established. 
This may also suggest a similarity of the dust fraction in different galaxies. Observations indicate no significant difference in the dust content over cosmic time for a given stellar mass and star formation rate \citep{Santini_et_2014}. 

The nuclear regions of ULIRG and SMG are characterised by extreme densities and high dust-to-gas ratios, leading to high IR optical depth. We have seen that in the IR-optically thick regime, the Eddington ratio does not depend on the column density and thus even dense material can potentially be disrupted. This seems to be a characteristic property of radiative feedback in optically thick environments \citep[][]{Murray_et_2010}. 
Moreover, enhanced dust-to-gas ratios also lead to higher Eddington ratios in the IR-optically thick regime, thereby facilitating the blow-out of dusty gas. 
Thus the more dusty medium, which is also responsible for heavy obscuration, is preferentially expelled by radiative feedback. 
Therefore the central AGN has a natural tendency to efficiently remove its own obscuring dust shroud and reveal itself. 

The actual importance of radiation trapping has been a source of much debate in the literature. 
It has been argued that the reprocessed IR photons tend to leak out through lower density channels, and that the rate of momentum transfer is largely over-estimated in spherical symmetry. For instance, \citet{Novak_et_2012} find that the momentum flux cannot reach values much exceeding $L/c$ (due to the low opacity of dust in the infrared and dust destruction processes) in simulations of early-type galaxies. On the other hand, 3D radiative transfer calculations, including multi-dimensional effects, indicate that outflows driven by radiation pressure on dust can attain momentum ratios of several times $L/c$ due to the multiple scatterings \citep{Roth_et_2012}. 
In general, the effects of radiation pressure are more important in environments characterised by high densities and large optical depths. We note that Hot DOGs are obscured by extreme column densities, consistent with Compton-thick \citep{Stern_et_2014, Piconcelli_et_2015}, such that the diffusion of reprocessed IR photons should have a significant impact on the ambient medium. 
Even if the radiation force is reduced by a factor of a few compared to the idealised case of a smooth spherical gas distribution \citep{Roth_et_2012, Skinner_Ostriker_2015}, the partial trapping of IR photons is still crucial in initiating the outflow. 

A further concern is the development of radiative Rayleigh-Taylor (RT) instabilities in the flow. Based on the flux-limited diffusion (FLD) approximation, \citet{Krumholz_Thompson_2013} find that the rate of momentum transfer is reduced due to the RT instabilities, and that the momentum ratio cannot considerably exceed the single scattering limit. On the other hand, using the variable Eddington tensor (VET) approximation, \citet{Davis_et_2014} show that the dusty gas can be continuously accelerated despite the development of the RT instabilities. 
A recent study comparing the different numerical methods suggests that the easy photon leakage seen in FLD simulations is partly due to the numerical approximation, and that there can be continuous acceleration of dusty gas in simulations using more accurate numerical schemes \citep{Tsang_Milosavljevic_2015}. Therefore, although some issues still need to be resolved, radiation pressure on dust remains a viable mechanism for driving large-scale outflows, especially in buried systems at early times. 

The large quantities of dust required in order to drive AGN feedback may be produced in the starburst phase. 
Recent observations indicate that substantial quantities of dust can be released in core-collapse supernovae \citep{Wesson_et_2015, Owen_Barlow_2015}. 
The resulting dust-to-gas mass ratio in the supernova remnants can be quite high, of the order of 1/30 in the case of the Crab Nebula, with a significant fraction of the dust grains surviving the shock due to their large sizes \citep{Owen_Barlow_2015}. 
It is well known that the dust opacity depends on the grain properties, such as size, structure, and composition. 
Coagulation and grain growth form larger aggregates, which are thought to be responsible for the increase in the dust opacity observed at long wavelengths \citep[][and references therein]{Kohler_et_2012, Ysard_et_2013}. 
Grain growth in the interstellar medium can also be an important factor that contributes to the dust mass increase at late times \citep{Michalowski_2015}. 
Large dust masses ($\sim 10^8-10^9 M_{\odot}$) are indeed observed in high-redshift SMG, which argue for supernovae with high yields, significant grain growth, and little dust destruction \citep[][and references therein]{Rowlands_et_2014}. 

Dust is produced when massive stars explode at the end of their lifetime, with typical timescales of the order of a few $\sim 10^6$yr. Thus a quite early build-up of dust may be expected in massive starbursts. 
Prolonged star formation in the nuclear starburst can keep the ambient medium dusty, which is an important requirement in our model, as the whole AGN feedback process relies on radiation pressure on dust. 
In this framework, AGN feedback and nuclear starbursts are naturally coupled through the production of dust in supernova explosions. While the connection between AGN and starbursts is often loosely attributed to a common source of fuel, we expect a natural causal link between the two phenomena. 
If dust production starts at a given redshift ($z_d$), then radiative feedback based on dust is only effective since $z < z_d$; but there are observational indications of dust formation in the early Universe \citep{Michalowski_2015}. 

Certainly, other forms of AGN feedback (jets and winds) are also operating at different stages in different galaxies. 
The relative importance of the radiative and kinetic feedback modes depends on the underlying accretion mode and varies with cosmic time. Quasar-mode feedback operates at accretion rates close to the Eddington limit, and is most effective at redshifts near the peak epoch of AGN activity ($z \sim 2$); while radio-mode feedback is usually associated with lower accretion rates and typically operates at lower redshifts \citep[][and references therein]{Fabian_2012}. In contrast to kinetic-mode feedback, which has strong observational evidence (e.g. in terms of radio bubbles and X-ray cavities), the radiative-mode is much more difficult to directly observe, due to obscuration. 
Nonetheless, evidence is starting to emerge for a number of sources, and much progress is expected with upcoming observations (e.g. ALMA). 

Following the removal of dusty gas by radiative feedback, the central source may appear as a bright quasar. 
This high-luminosity phase is however short-lived, and only lasts until the available fuel is consumed and the accretion disc is drained.  
The characteristic timescale can be quite short, of the order of $\sim 10^5$yr, for discs limited by self-gravity \citep{King_Pringle_2007}. Assuming that much of the accreting matter, i.e. the potential fuel supply, is swept away by the radiation pressure-driven outflow, the accretion rate falls off and the central source will eventually decline in luminosity.  
If observations happen to select sources after the AGN luminosity drop, the inferred value of the momentum ratio can be over-estimated \citep{Ishibashi_Fabian_2015}. 

We have seen that high momentum ratios are generally associated with high column densities in the obscured phase. 
However, if observations probe the later stages subsequent to the outburst event, one may observe high $\zeta$ values coupled with modest columns. 
In fact, a significant fraction of the sources in the observational samples are currently accreting at sub-Eddington rates \citep{Veilleux_et_2013, Cicone_et_2014}. 
The early phase of heavy obscuration, characterised by high momentum ratios and large columns, may actually be difficult to observe. 
A recent study indicates that the observed emission line ratios in quasar outflows favour driving by radiation pressure rather than pressure due to hot gas, and emphasises the importance of an earlier obscured phase \citep{Stern_et_2016}. 
In general, the physical conditions at the time when the outflow was initially launched may be quite different from the ones observed in local sources at the present time. 

\section*{Acknowledgements}

WI acknowledges support from the Swiss National Science Foundation and ACF acknowledges support from ERC Advanced Grant 340442.

  
\bibliographystyle{mn2e}
\bibliography{biblio.bib}


\label{lastpage}

\end{document}